# Growth of Ni and Ni-Cr Alloy Thin Films on MgO(001): Effect of Alloy Composition on Surface Morphology

Gopalakrishnan Ramalingam, Petra Reinke[a]

*Department of Materials Science and Engineering, University of Virginia, 395 McCormick Road, Charlottesville, Virginia 22904 USA*

The effects of substrate treatment, growth temperature and composition on the surface morphology of Ni-Cr thin films grown on MgO(001) are studied by scanning tunneling microscopy (STM) and atomic force microscopy (AFM). We demonstrate that a combination of acid-etched substrates and high temperature deposition (400 °C) will result in smooth films with well-defined terraces (up to 30 nm wide) that are suitable for the study of progression of chemical reactions on the surface. Two different treatments are used to prepare the MgO substrates for deposition and they introduce characteristic differences in film surface morphology. Thin films that are grown on the phosphoric acid-treated substrates present reduced nucleation density during the initial stages of film growth which results in long and wide terraces. Due to the ≈16% lattice mismatch in the Ni(001)/MgO(001) system, film growth at 400 °C yields discontinuous films and a two-step growth process is necessary to obtain a continuous layer. Ni films are deposited at 100 °C and subjected to a post-growth anneal at 300 °C for 2 hours to obtain a smoother surface. The addition of just 5 wt.% Cr drastically changes the film growth processes and yields continuous films at 400 °C without de-wetting in contrast to pure Ni films. With increasing Cr content, the films become progressively smoother with wider terraces. Ni5Cr alloy thin films have an rms surface roughness of 3.63±0.75 nm while Ni33Cr thin film is smoother with an rms roughness of only 0.29±0.13 nm. The changes in film growth initiated by alloying with Cr are due to changes in the interfacial chemistry which favorably alters the initial adsorption of the metal atoms on MgO surface and suggests a reduction of the Ehrlich-Schwoebel barrier. The growth of smooth Ni-Cr thin films with a well-defined surface structure opens up a new pathway for a wide range of surface science studies related to alloy performance.

---

[a] Author to whom correspondence should be addressed. Electronic mail: pr6e@virginia.edu

**Key words:** Thin films, Ni-Cr alloys, Epitaxy, Scanning tunneling microscopy, Ehrlich-Schwoebel barrier

## I. INTRODUCTION

Ni-Cr based superalloys have exceptional oxidation and corrosion resistance[1–6] and are currently used among other applications in nuclear power plants,[5] gas turbines,[7] and metallic dental implants.[8] Oxidation resistance of these alloys depends strongly on the Cr concentration and it has been found that addition of <8 wt.% leads to an increase in oxidation rate while alloying with 10-20 wt.% Cr results in reduced oxidation rates due to the formation of a continuous chromium oxide layer protects the material from further degradation.[9–11] Despite the technological relevance, the initial reaction steps in the oxidation of the alloy are not well understood. A facile route for fabrication of high quality NiCr alloy thin films would open up a path for the atomic scale study of oxidation as a function of Cr composition and prevents the need for cost intensive fabrication and cleaning of single crystals of different compositions. Majority of the work to investigate oxidation of NiCr alloys has been performed on bulk polycrystalline samples[10,12–15] where the effect of Cr concentration cannot be decoupled from other factors such as grain size, residual stresses and crystallographic orientations. In contrast, the use of smooth alloy thin films provides a direct route to study the initial stages of alloy oxidation as a function of Cr content with high spatial resolution.

The growth temperature and composition for the growth of smooth Ni-Cr alloy thin films should be such that we are operating in the solid solution regime. The phase diagram shows that Cr forms a substitutional solid solution with Ni but the solubility limit of Cr can only be extrapolated for T<500 °C from current data.[16] Experimental data exists only for high temperatures and calculated phase diagrams differ in the Cr solubility limit as well as in predicting the presence of the $Ni_2Cr$ intermediate compound. In addition, the lattice constant of the solid solution of Cr in Ni needs to be considered to determine the epitaxial constraints during growth. The constituent elements Ni and Cr have different crystal structures: Ni crystallizes in a face centered cubic (FCC) structure with lattice constant, $a_o$=0.352 nm while Cr crystallizes in the body centered cubic (BCC) structure with $a_o$=0.288 nm.[17] The lattice mismatch between Ni(001) and MgO(001) at room temperature is ≈16% while the mismatch between Cr(001) and MgO(001) is only 3.4%; in case of Cr, the unit cell is rotated by 45° such that the epitaxial relationship is (001)Cr||(001)MgO, [110]Cr||[100]MgO. The change in lattice parameter due to Cr dissolution in Ni is negligible and increases by only 0.004 nm after alloying 40 wt.%Cr with Ni.[18,19] Consequently, the growth of Ni-Cr alloy thin films should follow the characteristics of Ni thin film growth and changes in the thin film structure can be attributed to chemical effects such as modified adsorption of atoms on MgO surface or changes in surface mobility in the presence of Cr.

The growth of Ni thin films on MgO has been studied extensively and the film structure, in terms of surface morphology and terrace structure, was found to strongly depend on the growth parameters such as temperature and film thickness.[20–24] Two dominant epitaxial orientation relationships are established: (001)Ni||(001)MgO,

[100]Ni‖[100]MgO and (1$\bar{1}$0)Ni‖(001)MgO, [11$\bar{2}$]Ni‖[100]MgO.[21,23,24] The epitaxial relationship is influenced by the film thickness[25] and thinner films (25-40 nm), such as the ones used in our study, favor the (100) cube-on-cube orientation. It should be noted that orientation relationships depend on growth rates (deposition flux) and the literature cited here use different deposition fluxes compared to the current work. The growth temperature also affects the crystallographic orientation of the Ni thin film and Ni(100) cube-on-cube orientation has been reported[20,21] for deposition at 100 °C while the (1$\bar{1}$0)Ni orientation is found when films are grown at 350 °C.[23] Recently, a two-step process involving deposition at 100 °C and annealing at 300 °C was reported[20] to obtain smooth (100)-oriented films with well-defined terrace structure.

In this report, we demonstrate the growth of smooth $Ni_{1-x}Cr_x$ thin films (0≤x≤35wt.%) with terrace widths of up to 30 nm on MgO(001) substrates and discuss the critical aspects of the growth process. The quality of the films as expressed by the terrace width, surface morphology and roughness are characterized by scanning tunneling microscopy (STM) and atomic force microscopy (AFM). The NiCr alloy thin film growth depends sensitively on (i) surface structure of the MgO substrate (terrace sizes and step edge structure), (ii) growth temperature (100 °C and 400 °C) and (iii) Cr content in the alloy (0-35 wt.%).

## II. EXPERIMENTAL DETAILS

The experiments are performed in an Omicron Nanotechnology Variable Temperature Scanning Probe Microscopy (VT-SPM) system under ultrahigh vacuum (UHV) conditions. The base pressure is <3×10$^{-10}$ mbar. Epi-polished MgO(100) crystals (CrysTec GmbH) are used as substrates for thin film growth. Table I summarizes the substrate pre-treatments and growth conditions that are used in the current work. Based on the existing literature for metal film growth on MgO substrates,[20,26–28] two different pre-treatment routes are used to prepare a smooth substrate surface for film growth: (T1) annealing at 600 °C in-situ (UHV conditions) for 2 hours immediately following an overnight anneal at 200 °C and (T2) two-step acid etch plus anneal route: the substrate is etched with 85% phosphoric acid for 30 seconds and annealed in air in a box furnace at 1200 °C for 5 hours followed by a second 30 second etch and annealing in air at 1100 °C for 4 hours; this is followed by overnight anneal after introduction into the UHV chamber. The etching process removes the rough $Mg(OH)_2$ structures from the surface while the high temperature anneal in air recovers the crystalline surface as shown in Figure S1 in the supplemental information. The overnight anneal at 200 °C in the UHV chamber is used to remove adventitious hydrocarbons and water. Ni (Alfa Aesar, 99.999% purity) and Cr (American Elements, 99.95% purity) are deposited by electron beam evaporation using a Mantis EV mini e-beam evaporator (QUAD-EV-C). The deposition rates of Ni and Cr are measured using a quartz crystal monitor and the deposition time is estimated to achieve a film thickness of 30 nm. The Ni flux is maintained between 0.30-0.35 nm/min and the Cr flux (0.01-0.14 nm/min) is adjusted to obtain the required composition of the alloy. Two growth conditions are used in this study: (G1) deposition at 100 °C followed by annealing at 300 °C for 2 hours, and (G2) deposition at 400 °C.

    The as-grown thin films are characterized in situ by scanning tunneling microscopy. Imaging is performed at room temperature in constant current mode with W tips which are prepared by electrochemical etching. Empty and filled states images are

recorded on the different films; low bias voltages ($V_b$<0.2 V) and set-point currents greater than 0.5 nA resulted in good quality images. Surface morphology of the pre-treated MgO substrates are characterized ex-situ using atomic force microscopy (AFM) with an NT-MDT Solver Pro in tapping mode, using ETALON HA/NC tips with curvature radius less than 10 nm. The composition of the film is confirmed ex-situ by energy dispersive spectroscopy (EDS) performed in an FEI Quanta 650 scanning electron microscope (SEM). Film thickness estimates are confirmed by X-ray reflectometry (XRR) on films grown on T1-substrates while the rough film-substrate interface in the case of T2 films precluded reliable XRR measurement (see Figure 1 for MgO surface morphology). Time of flight secondary ion mass spectroscopy (ToF-SIMS) was performed on two randomly chosen films to check for uniformity of alloy composition. ToF-SIMS measurements were performed on an ION-TOF (Gmbh) instrument at Surface Science Western (University of Western Ontario, Canada) and detailed information is provided in the Supplemental Information. STM and AFM images were analyzed with Gwyddion, an open source software for SPM data analysis.[12] A blue color scheme is chosen for all AFM images and an orange scheme is used for all STM images shown in this work.

Table I. List of alloy composition with substrate pre-treatment and growth conditions.

|  | T1 Pre-treatment<br>600 °C UHV Anneal | T2 Pre-treatment<br>H3PO4 etch + Air Annealing |
|---|---|---|
| G1<br>100 °C Growth +<br>300 °C Anneal | Ni, Ni-13Cr, Ni-23Cr | Ni |
| G2<br>400 °C Growth | Ni | Ni, Ni-5Cr, Ni-13Cr, Ni-14Cr, Ni-33Cr |

## III. RESULTS

### A. Effect of MgO Pre-Treatment

The AFM images of the MgO surface after T1 and T2 pre-treatment are shown in Figure 1(a,b). The MgO surface after T1 pre-treatment is smooth with an average RMS roughness of 0.18±0.04 nm but no terrace structure is observed. The T2 pre-treatment yields a terrace structure with well-defined step-edges. The average RMS roughness of T2 substrates is 2.07±0.47 nm. The higher roughness results from the residual features left behind by the etching of Ca spires (shown in Figure S1(c)) and the stacked terraces that result from high temperature annealing. AFM images of the as-received MgO surface and those recorded at the intermediate stages of the T2 pre-treatment are shown in Figure S1 in supplemental information. The higher temperature annealing used in the T2 pre-treatment results in wide terraces (80-150 nm) and the step edges are predominantly oriented along the <110> direction. The T2 pre-treated substrates retain their terrace structure for at least 3 weeks when stored in a vacuum dessicator and yield good quality thin films.

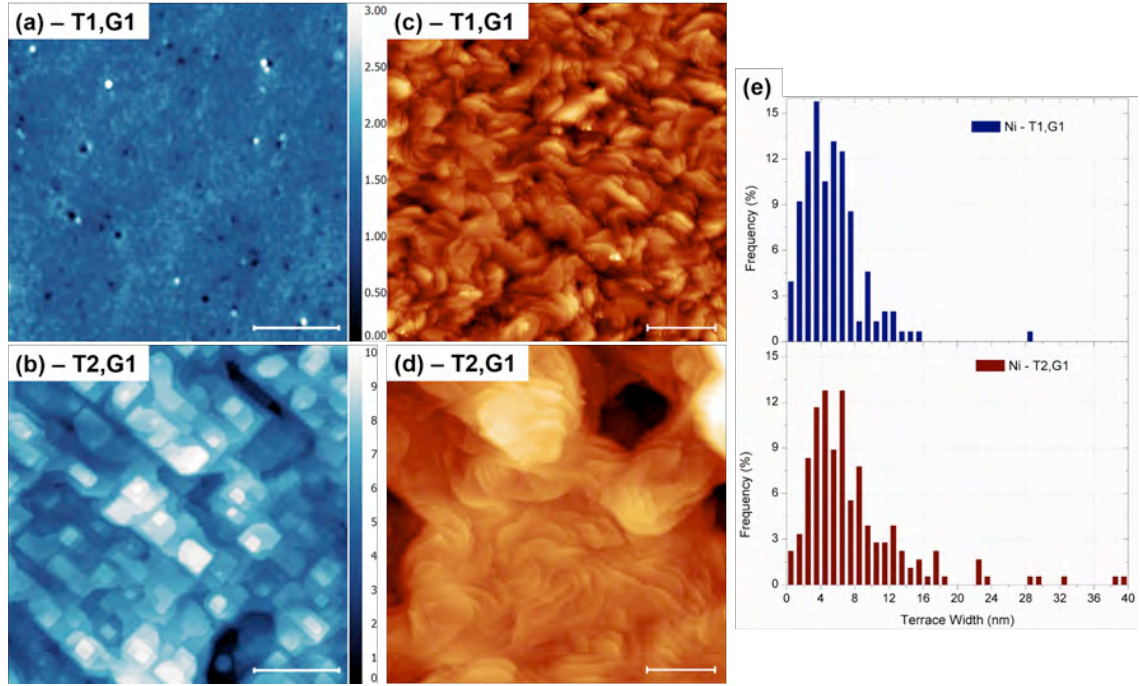

**Figure 1.** AFM images of MgO substrates after (a) T1 and (b) T2 pre-treatments (scale bar is 500 nm). STM images of Ni thin film grown on (c) T1 and (d) T2 substrates using the G1 growth recipe (scale bar is 50 nm). Color scale in AFM images is in nm.

STM images of the Ni thin films grown using G1 conditions on T1 and T2 substrates are included in Figure 1(c) and (d). The films show well-defined terraces on both the substrates. The average terrace width, measured along a line perpendicular to the step edges, is 5.26±3.6 nm and 7.85±6.3 nm on T1 and T2 substrates respectively. The large spread in the terrace width on T2 substrates is due to a few instances of very wide terraces (>25 nm) and the terrace width distribution is shown in Figure 1(e). The terrace boundaries in Figure 1(d) are rounded and very few straight step edges are present. There is a long-range roughness associated with the etching-induced inhomogeneity for films on T2 substrates including the depression seen in Figure 1(d). Despite the similar terrace widths and step heights for Ni thin films grown on T1 and T2 substrates, the two surfaces look distinctly different. The terrace density is significantly lower on the T2 substrates and the terraces of films on T2 substrates are longer (measured parallel to the step edge). For identical terrace widths, long terraces observed on Ni films grown on T2 substrates results in a greater aspect ratio of the terraces when compared to films on T1 substrates. The lack of a texture on the T1 substrates results in explosive nucleation of Ni clusters while the nucleation density and consequently the terrace density is lower on crystalline T2 substrates. Therefore, the T2-MgO substrates that suppress the nucleation density during initial stages of film growth and yield thin films with long and wide terraces are recommended for good quality thin film growth.

### B.  Effect of Growth Temperature

AFM and STM images of Ni thin films grown using G1 and G2 conditions on T2 substrates are shown in Figure 2. The Ni thin film deposited at 100 °C (step 1 of G1 growth method) is rough with small islands and lacks the well-defined terrace structure (Figure 2(a)). The rms surface roughness of this surface is 0.37±0.05 nm and the morphology is consistent with literature.[20] Subsequent annealing at 300 °C for 2 hours (step 2 of G1 method) leads to a smoother film (Figure 2(b)) and was discussed in detail in the previous section with reference to Figure 1(d,e). We grew Ni thin films (30-60 nm) using the G2 conditions but did not achieve tunneling contact for STM imaging since the films are still discontinuous and below the percolation threshold for conduction.

Ex-situ AFM measurements of Ni films shown in Figure 2(c) and 2(d) illustrate the differences in the long-range surface morphology of G1 and G2 depositions. As expected from STM observations, the G1-Ni films are smooth with rms roughness of 2.7±0.4 nm similar to a T2-treated MgO surface and the surface features are 3-8 nm in height. On the other hand, the G2-Ni film surface shown in Figure 2(d) feature an island-type morphology which is favored at higher temperatures. The high Ni/MgO interfacial energy (due to the 16% lattice mismatch which results in an incoherent interface) and increased surface mobility at 400 °C promotes island growth and leads to poor wetting to minimize contact area. The rms roughness of this surface is 9.3±0.9 nm and the islands heights are in the 30-50 nm range with some islands exceeding 80 nm. The surface shown in Figure 2(d) is not sufficiently conductive to be measured with STM.

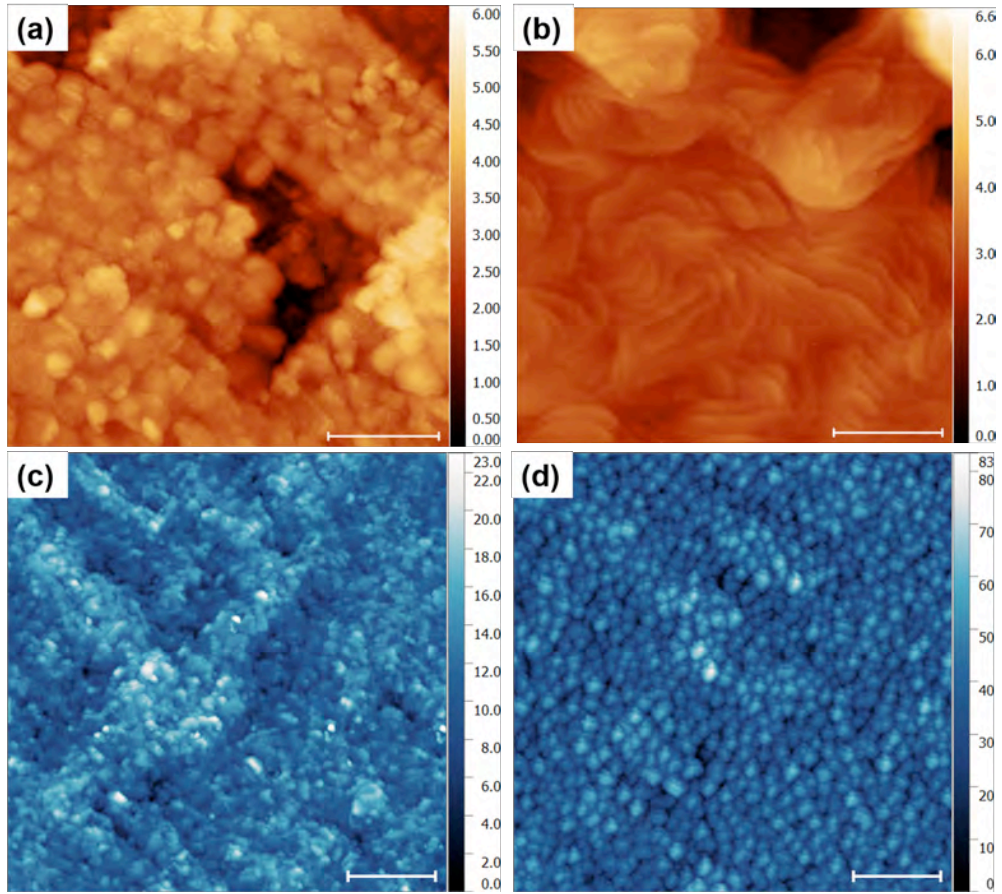

**Figure 2.** STM and AFM images show the effect of growth temperature on the Ni thin film morphology. STM images after (a) deposition at 100 °C (step 1 of the G1) and (b) post-deposition annealing at 300 °C for 2 hours (step 2 of G1) – same deposition as shown in Figure 1(d); scale bar is 50 nm. Ex-situ AFM images of Ni film grown using (c) G1 and (d) G2 growth conditions; scale bar is 1 μm. Color scale is in nm.

STM images of alloy thin films grown using G1 and G2 conditions are shown in Figure 3. The G2-type growth of the Ni-13Cr alloy yields a surface with crystalline terraces as shown in Figure 3(a). However, the alloy surface is interspersed with small 3D structures and these are likely intermediate compounds in the Ni-Cr system such as $Ni_2Cr$. The terrace sizes (shown in a histogram in Figure 3(c)) are consequently small due to the presence of these 3D structures and have an average width of only 4.78 nm with a spread of 2.62 nm. Figure 3(b) shows the Ni-13Cr film grown using G2 conditions: in contrast to pure Ni thin films, the G2 growth condition yields continuous thin films in the alloy which are sufficiently conducting to be characterized with STM. The average terrace width is 11.57±8.86 nm; the large width of the distribution is due to the numerous narrow terraces owing to the high density of step edges at the perimeter of the terraces. 40% of the terraces are less than 8 nm wide while the remaining 60% of the terraces are 10-30 nm wide. The terraces in alloy thin films are significantly larger than those obtained in Ni films and the combination of (T2,G2) is ideal for growth of good quality alloy thin films. At higher temperatures, we operate in the solid solution region of the Ni-Cr phase diagram thereby preventing the formation of intermediate compounds. Higher surface mobility due to the higher deposition temperature explains the smoothness of alloy films grown in G2 conditions. However, the contrast between Ni-Cr alloy films and the Ni films using the G2 growth route and the wider terraces obtained in alloy thin film warrants a deeper study of the effect of Cr on the growth behavior.

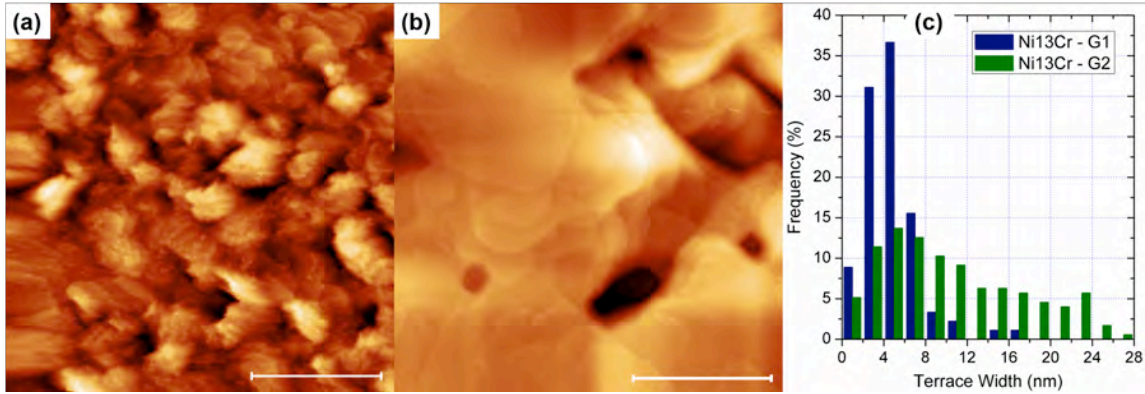

**Figure 3.** STM image of Ni13Cr alloy thin film showing the effect of growth temperature on surface morphology: (a) (G1,T1) conditions and (b) (G2,T2) conditions. Scale bar is 50 nm. (c) Terrace width distribution of Ni-13Cr thin film grown using G1 and G2 conditions.

## C. Effect of Cr Alloying

In this section, we discuss the impact of increasing Cr concentration on the film terrace structure and roughness. For the same film thickness of ≈30 nm, Ni-Cr alloy thin films grown using G2 conditions on T2 substrates are continuous and electrically conductive and the full complement of STM and AFM images is available. The addition of just 5 wt.% of Cr (lowest in this study) is sufficient to alter the growth and yields continuous films amenable for STM characterization as shown in Figure 4(a). The derivative image (current image) is shown in this case due to extreme height variations. The film is very rough and is characterized by the coalescence of large islands. These islands have a flat top with well-defined terraces and a high density of step-edges at the island perimeters. With increasing Cr concentration, the films progressively smoothen as shown in Figure 4(b) and (c). This smoothening is represented in the variation of rms roughness of the films with increasing Cr concentration: Ni5Cr, Ni14Cr, and Ni33Cr have a roughness of 3.63±0.75, 0.86±0.04, and 0.29±0.13 nm respectively. The island in Ni5Cr image shown in Figure 4(a) is ≈11 nm and the overall island height decreases with increasing Cr content with islands only 6 nm and 2 nm in case of Ni14Cr (Figure 4(b)) and Ni33Cr (Figure 4(c)) respectively. The terrace structure also changes with increasing Cr concentration and straighter step edges are observed indicative of changes in the step edge energies with increasing Cr content. We hypothesize that the addition of Cr to Ni alters the step edge energies so that it is energetically favorable for the system to decrease step edge length (straighter edges). The terrace structure of Ni33Cr closely resembles the surface of Cr thin film grown on MgO shown in Figure S2. The large-area scan of the Cr surface is characterized by predominantly straight step edges and the surface resembles the initial MgO terrace structure (Figure 1(b)). However, high resolution images confirm the presence of some curved step edge segments indicative of some of the similar growth dynamics that are observed in the Ni-Cr alloy films. It should be noted that the distribution of Cr throughout the alloy film is uniform as seen in the ToF-SIMS results obtained from a Ni13Cr film (Figure S3 in supplemental information). No significant Cr segregation is observed at the MgO-alloy interface and the Cr and Ni ion yields show a similar decrease at the film interface. Therefore, the changes in film growth with Cr additions are not due to preferential wetting of MgO by Cr atoms.

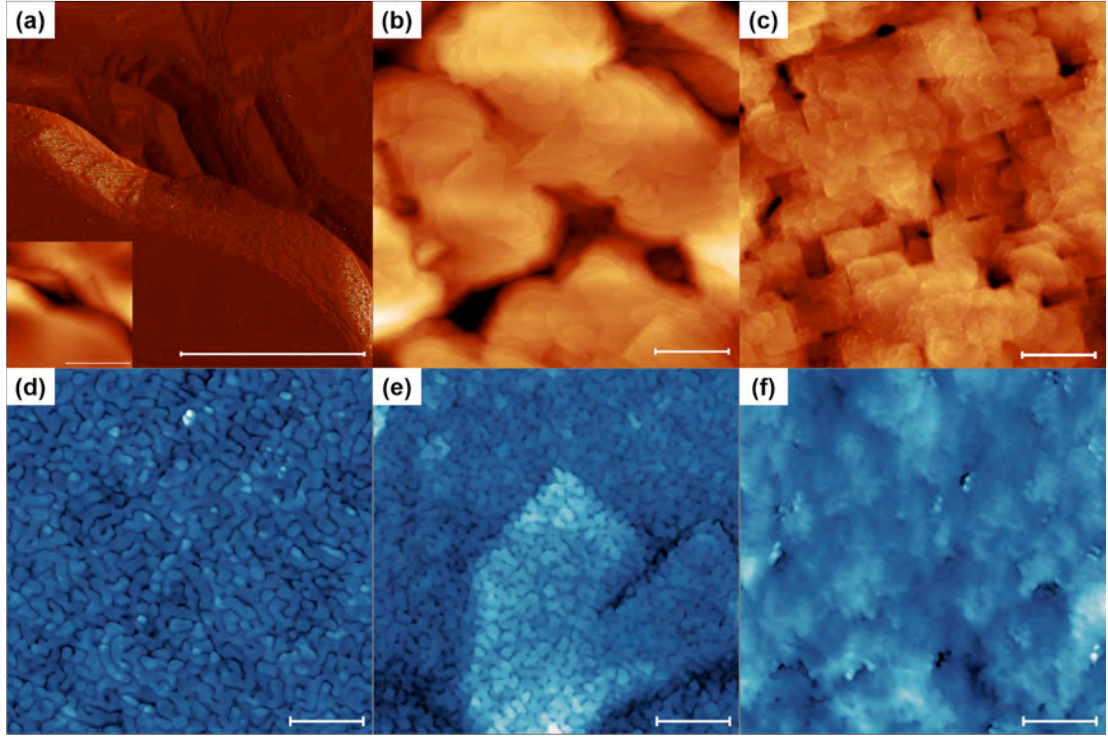

**Figure 4.** STM and AFM images of alloy thin films showing the effect of Cr addition on surface morphology: (a,d) Ni5Cr, (b,e) Ni14Cr and (c,f) Ni33Cr films grown on T2 substrates using G2 growth recipe. Scale in STM and AFM images is 50 nm and 1 μm respectively. The derivative STM image (current image) is shown in (a) due to large height variations in the image; the inset shows the topographic image.

AFM images of the Ni5Cr, Ni14Cr, and Ni33Cr alloy films are shown in Figure 4(d-f). Consistent with the observations in STM, roughness of the film decreases with increasing Cr concentration. In addition, the island stacks prevalent in lower Cr content alloys are almost absent in the Ni33Cr alloy. The variation of island stack heights measured from AFM images as a function of Cr concentration is shown in Figure 5. The height of an island stacks is measured from AFM images except for Ni33Cr film where the island height is obtained from STM image. The alloy films are markedly different from Ni film (shown in Figure 2(d)) with the former characterized by wider but continuous islands while the latter consists of smaller but discontinuous islands. The average width of the island stacks in Ni5Cr and Ni14Cr films is nearly identical measured as 119.6±24.47 nm and 114.24±18.33 nm respectively. Ni islands are smaller with an average width of 104.34±16.31 nm. The average island stack height decreases with increasing Cr content in the alloy: the addition of only 5wt.% Cr to the matrix results in a 70% reduction in the island height and the decrease continues to the point where individual islands are barely distinguished in the AFM images of the Ni33Cr film. From these findings, we can conclude that the addition of Cr to the Ni matrix changes the initial wetting of the MgO and the subsequent film growth. The reduced height of the islands with increasing Cr content also suggests that interlayer diffusion is improved. An important kinetic factor that is likely affected by the Cr addition is the Ehrlich-Schwoebel (E-S) barrier.[30–32] With increasing Cr content of the alloy, the E-S barrier decreases which results in decay of large islands as atoms can diffuse across step edges and leads to an overall flattening of islands as seen in the images in Figure 4 and the plot in Figure 5.

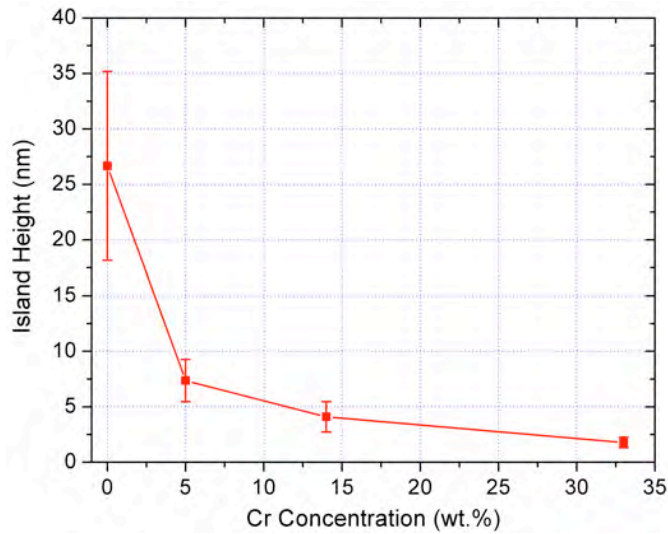

**Figure 5.** Variation of island heights in alloy thin films as a function of Cr concentration. The island heights are measured from AFM images except for the Ni33Cr sample where STM images are used as the individual islands are not distinguished in AFM images.

## IV. DISCUSSION

The growth of smooth Ni and NiCr thin films on MgO(001) was demonstrated and we will now discuss the important factors that affect the initial nucleation and growth process which manifests itself as a change in the terrace widths and/or surface morphology of the films.

The preparation of the MgO surface has a significant impact on the initial nucleation density of islands and affects the surface structure of the films as shown in Figure 1. The terraces of Ni film grown on T2 substrates are longer compared to films grown on T1 substrates and consequently have a greater aspect ratio (length/width ratio). The T1-MgO surface promotes greater nucleation density, which leads to the smaller terrace/island sizes observed. This is a common observation on "glassy" substrates which are relatively smooth but possess chemical disorder and nanoscale inhomogeneities facilitating heterogeneous nucleation. The T2 surface which shows very well-defined crystalline ordering and the cube-type terrace structure typical for MgO promotes the formation of terraces in the metal overlayers. The T2 preparation favors a smaller nuclei density and thus larger crystallites, which can be concluded from the comparison of Figures 1(c) and 1(d). High temperature deposition (G2) leads to de-wetting of the Ni thin film and is independent of the MgO substrate pre-treatment. A discontinuous thin film even on the glassy T1-substrate suggests that the de-wetting is a consequence of the high interfacial energy between film and substrate. As a consequence of the de-wetting, the electrical conduction is reduced and thus contact is insufficient for STM measurements.

The surface morphology also exhibits distinct differences in terrace shape and spatial arrangement when comparing G1 and G2 growth modes, and an even larger variability as a function of Cr-concentration. A helical or screw-like arrangement of the terraces is seen in Ni-depositions using T2-G1 and T2-G2 and in most alloy thin films, but not in the low-T first step of T2-G1 (Figure 2(a)). The helical screw-like pattern has its origin in screw dislocations that are propagated during film growth. Possible sources of dislocation include inheritance from the substrate due to the T2 etch and anneal process, incoherent meeting of growth fronts and at the interface during growth to relieve mismatch strain.[33–37] The terrace width is controlled by the pitch of the dislocation and remains fairly constant for all experiments discussed here (Figure 1-4). The Ni/NiCr islands grow outwards from the dislocation core lengthening and widening the terrace as the process continues. The length of the terrace on the other hand increases with distance from the dislocation core and has been observed in other systems where screw-dislocation mediated film growth prevails.[33,34,38] A higher growth temperature such as the 400 °C used in G2 method will lead to a widening to the terrace farther from the core as seen in the terrace width distribution in Figure 3(c) which shows a greater terrace width for alloy films grown at higher temperatures. The long-range helical arrangement of the terraces is mostly retained and is identical for all alloy experiments. The large misfit at the (001)Ni||(001)MgO interface can lead to recrystallization and local reorientation of the Ni crystallites in order to relieve the strain energy. However, such a recrystallization is only reported for growth temperatures (400 ºC for G2) for a critical film thicknesses exceeding 50 nm which is outside the range of our experiments.[25] A slight reduction in the interfacial misfit with Cr alloying[18] is expected to further increase the critical film thickness and recrystallization is not expected to have a significant impact on the alloy growth. The strong impact of Cr concentration on the film morphology is, therefore, attributable to the atomistic processes of growth and the interfacial chemistry rather than considerations of epitaxy as observed in the growth of $Cr_xV_{1-x}$ thin films on MgO(001).[39]

It has been shown that Ni strongly interacts with the oxygen ions on the MgO surface and preferentially forms a covalent polar bond at the anionic sites while Cr only has a weak interaction on anionic sites of MgO surface.[40–42] In any given alloy deposition experiment, the anionic adsorption sites are expected to be occupied by Ni. In this scenario, we posit that the weakly interacting Cr atoms adsorb on the nearby cationic site (occupying up to ≈14% of the sites in Ni14Cr). The presence of Cr on this site will statistically hinder the Ni occupation on anionic sites and increase the Ni-Ni distance towards equilibrium lattice spacing locally reducing strain and thereby leading to local stability. These local chemical changes on the surface due to the presence of Cr are suggested as a model to explain the improved wetting of the MgO substrate during the initial stages of growth of alloy thin films which can consequently lead to smoother films.

## V. CONCLUSIONS

We have demonstrated the growth of smooth Ni(001) and Ni-Cr(001) alloy thin films with wide, well-defined terraces on MgO(001). High temperature deposition at 400 °C on acid-treated substrates is recommended for obtaining the best quality Ni-Cr alloy thin films. The growth of thin films in this system is mediated by screw dislocation at the film-substrate interface which leads to the characteristic helical pattern of the terraces. Increasing Cr concentration in the alloy leads to a decrease in Ehrlich-Schwoebel barrier as manifested by progressively smoother films with large terraces. The drastic reduction in surface roughness and changes in film growth mechanism with Cr alloying are mainly driven by changes in the interfacial chemistry. The study of NiCr system is uniquely suited to further our understanding of thin film growth. The near constant lattice mismatch between NiCr-MgO up to 35wt% Cr enables us to decouple the impact of interfacial reactions, and local energy barriers such as the Ehrlich Schwoebel barrier, and the role of the strain field mitigated by lattice mismatch. In addition, such a method for the growth of thin films of Ni-Cr alloys, which are recognized for their exceptional corrosion resistance, opens up a new pathway for the study of reactivity of these alloys as a function of alloy composition.

**SUPPLEMENTARY MATERIAL**

See supplementary material for (a) AFM images of the MgO surface during various stages of T2 pre-treatment, (b) STM images of Cr thin film, and (c) ToF-SIMS results of Ni13Cr thin film.

**ACKNOWLEDGMENTS**

This work is supported by ONR MURI "Understanding Atomic Scale Structure in Four Dimensions to Design and Control Corrosion Resistant Alloys" on Grant Number N00014-14-1-0675. The authors acknowledge Prof. Jerrold A. Floro for providing access to the AFM instrument and Surface Science Western at University of Western Ontario for performing ToF-SIMS measurements on the NiCr thin films.

[42] I. Yudanov, G. Pacchioni, K. Neyman, and N. Rösch, J. Phys. Chem. B **101**, 2786 (1997).

# Growth of Ni and Ni-Cr Alloy Thin Films on MgO(001): Effect of Alloy Composition on Surface Morphology

Gopalakrishnan Ramalingam, Petra Reinke[a)]

University of Virginia, Department of Materials Science and Engineering, 395 McCormick Road, Charlottesville, Virginia 22904 USA

## SUPPLEMENTARY INFORMATION

**MgO Substrate Pre-treatment**

AFM images of the MgO substrate after various stages of T2 pre-treatment[1,2] are shown in Figure S1. The as-received sample is rough and is contaminated by adsorbates such as water and hydrocarbons from ambient atmosphere; the structures in Figure S1(a) are 30-60 nm high while some larger impurity particles are also seen. After the first $H_3PO_4$ etch, the surface is more homogeneous but is still corrugated as seen in Figure S1(b); the surface structures are less than 20 nm high. After annealing in air at 1200 °C, the surface is populated with large $CaCO_3$ spire structures that are formed due to Ca segregation to the surface.[3] The AFM image in Figure S1(c) shows the presence of these spires on an otherwise smooth surface. Terraces and step edges are observed after this 1200 °C annealing step. The terraces are at least 100 nm wide with well-defined step-edges; step bunching and pinning due to the $CaCO_3$ structures are also clearly observed. The inset in Figure S1(c) shows the population of Ca spires throughout the surface (inset area is 10x10 $\mu m^2$). The second phosphoric acid etching and subsequent 1100 °C anneal yields a smooth surface as shown in Figure 1(d) without any calcium carbonate spires: the lower annealing temperature precludes any further Ca segregation and the long-range surface diffusion at these temperatures results in a smooth terrace structure suitable for thin film deposition. Line profiles shown in Figure S1(e) captures the large Ca spire structures as well as the flat terrace structure after the 2nd etch.

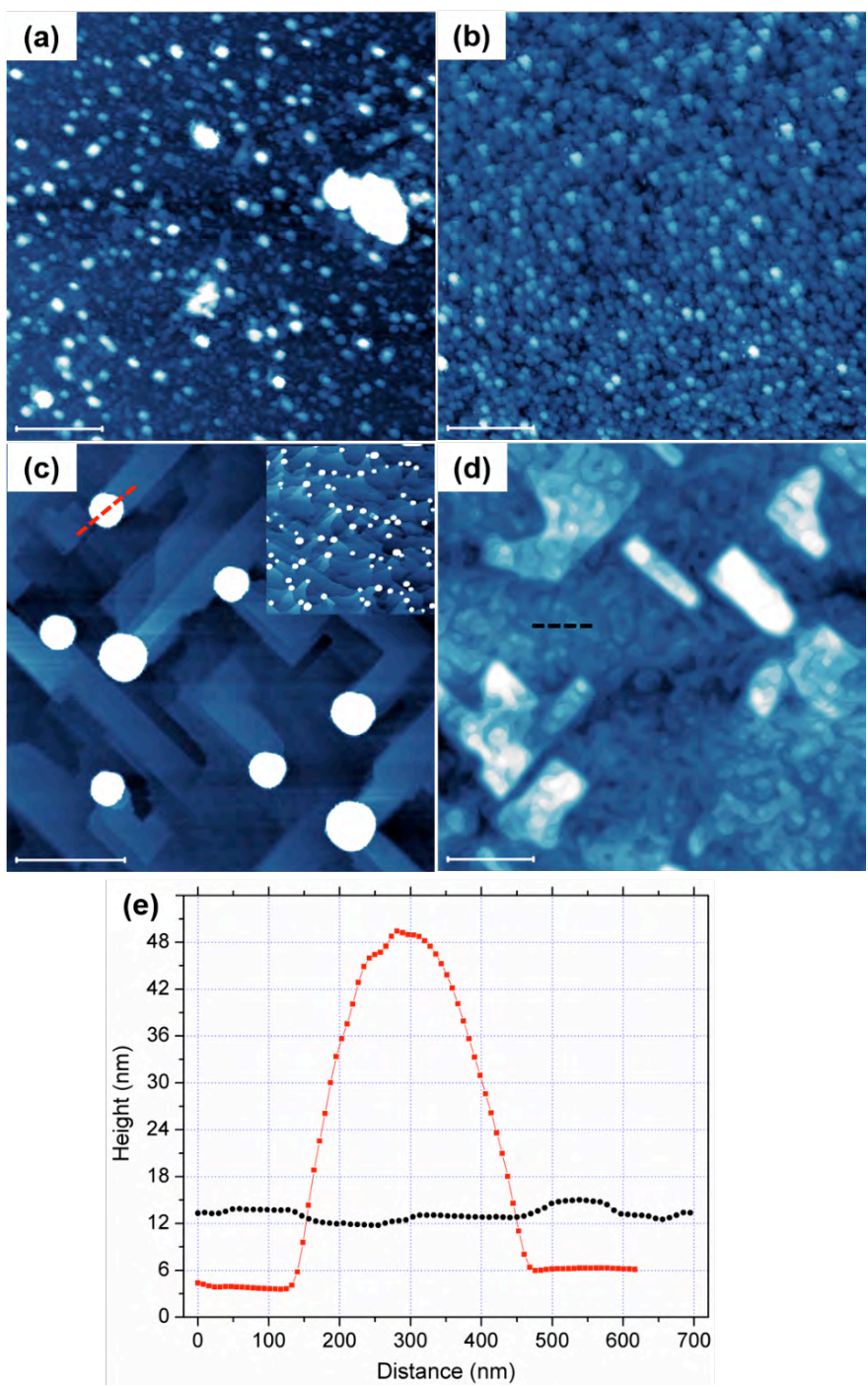

**Figure S1.** AFM images during different stages of MgO pre-treatment: (a) as-received, (b) after $H_3PO_4$ etch, (c) post-1200 °C annealing in air, and (d) after second acid etch and 1100 °C anneal. The inset in (c) is a 10×10 $\mu m^2$ area showing the distribution of Ca spires

throughout the surface. (e) Height profile on the surface across red and black lines marked in (c) and (d) respectively. Scale bar in all images is 1 $\mu$m.

**Cr Thin Film Growth**

STM images of Cr thin film grown using the recipe used by Wang et al.[4] are shown in Figure S2. After deposition at room temperature (with a Cr flux of 0.29 nm/min), the surface is rough (Figure S2(a)) with no evident terrace structure except for the faint imprint of the underlying MgO substrate After annealing at 550 °C, the film smoothens and has well-defined terraces (Figure S2(b)). The Cr film has terraces with predominantly straight step edges indicative of good epitaxial matching with the MgO surface (Figure 1(b)). The epitaxial relationship for the Cr/MgO(001) system is (001)Cr∥(001)MgO, <110>Cr∥<100>MgO. High resolution images of the surface are shown in Figure S2(c,d) and shows some circular step edges that are not resolved in the large area image shown in Figure S2(b). The atomic rows in Figure S2(c) which are along the <110> directions on the Cr(001) plane are oriented along the MgO<100> corresponding to the <110>Cr∥<100>MgO epitaxial matching that is expected in this system. The small segments in Figure S2(d) which present a different atomic structure are likely due to a trace adsorbate-induced reconstruction – various adsorbate-induced reconstructions have been discussed in literature[5–7] but none of these match the structures observed here.

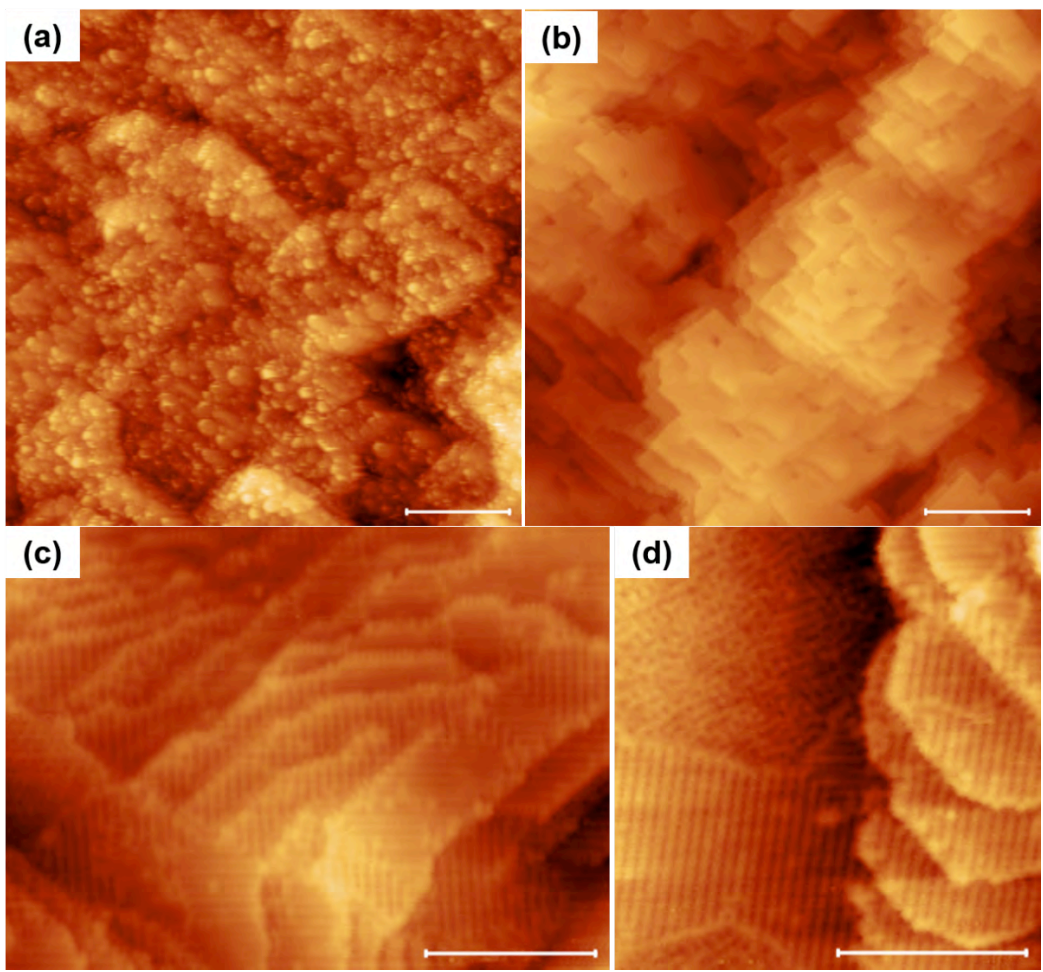

**Figure S2.** Large-area STM images of the Cr thin film grown on MgO(001): (a) as-deposited at room temperature, and (b) post-deposition anneal at 550 °C for 2 hours. Scale bar is 100 nm. High resolution images of the surface after 550 °C anneal showing (a) the atomic rows of the Cr film along <110>, and (b) small pockets of trace adsorbate-induced reconstruction on an otherwise clean surface. Scale bar is 10 nm in (c) and (d).

**Time of Flight Secondary Ion Mass Spectroscopy (ToF-SIMS) on Ni13Cr**

The measurements were performed using an ION-TOF (GmbH) TOF-SIMS IV equipped with a Bi cluster liquid metal ion source at Surface Science Western (University of Western Ontario, Canada). A 25 keV $Bi_3^+$ cluster primary ion beam pulsed at 10 kHz was used to irradiate the sample surface to generate secondary ions. In order to profile through the film thickness, a 10 keV $C_{60}^+$ ion beam was used to sputter the surface in an area of 300×300 μm$^2$ and positive ion mass spectra were collected. Depth was calibrated by measuring the craters generated using a stylus profiler.

From the data shown in Figure S3, we see that the alloy composition is uniform throughout the film. Cr and Ni ion yields ratios are identical throughout the film and begin to drop at the same depth indicating no compositional segregation in the film. Since the ion yields of both elements vary differently as a function of their chemical environment, we cannot make quantitative conclusions based on ion yields. However, it is clear that Ni and Cr ion yields start to drop at the same depth (≈27 nm) and follow the same trend and plateau at ≈37 nm – this confirms, within the resolution of the instrument, that there is no segregation at the interface. The Ni yield at the surface is reduced (Cr segregation at the surface) and is likely due to the formation of a passive chromium oxide layer on the surface of the Ni-Cr alloy film that serves to protect the film from further oxidation.

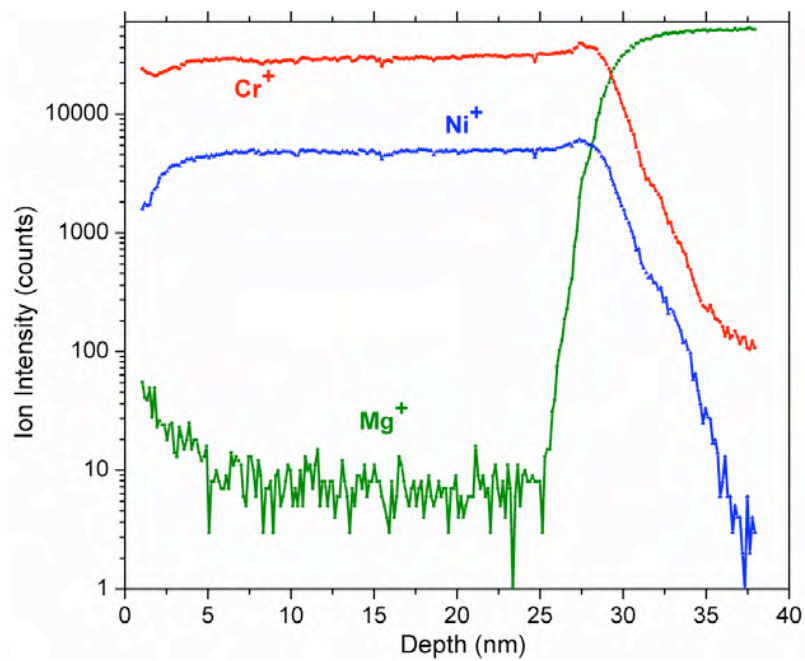

**Figure S3.** ToF-SIMS data of the Ni13Cr sample grown on T2-MgO indicating uniform film composition and no segregation at the film-substrate interface.